**A nutritional strategy to promote gilthead seabream performance under low temperatures**

Rita Teodósio[a,b], Cláudia Aragão[a,b], Rita Colen[a], Raquel Carrilho[a], Jorge Dias[c], Sofia Engrola[a*]

[a] *Centre of Marine Sciences (CCMAR), Universidade do Algarve, Campus de Gambelas, 8005-139 Faro, Portugal*

[b] *Universidade do Algarve, 8005-139 Faro, Portugal*

[c] *SPAROS Lda., 8700-221 Olhão, Portugal*

[*]Corresponding author:

Sofia Engrola

E-mail address: sengrola@ualg.pt

Centre of Marine Sciences of Algarve (CCMAR)

Universidade do Algarve,

Campus de Gambelas, Building 7

8005-139 Faro, Portugal

Tel: +351 289 800 100 x 7370

## Abstract

Gilthead seabream (*Sparus aurata*) is vulnerable to low water temperature, which may occur in the Southern Europe and Mediterranean region during Winter. Fish are poikilothermic animals, therefore feed intake, digestion, metabolism and ultimately growth are affected by water temperature. This study aimed to evaluate growth performance, feed utilisation, nutrient apparent digestibility, and nitrogen losses to the environment in gilthead seabream juveniles reared under low temperature (~13°C). Three isolipidic and isoenergetic diets were formulated: a diet similar to a commercial feed (COM) that contained 44% crude protein and 27.5% fishmeal, and two experimental diets with a lower protein content of 42% (ECO and ECOSup). In both ECO diets fishmeal inclusion was reduced (10% in ECO and 7.5% in ECOSup diet) and 15% poultry meal was included. Additionally, the ECOSup diet was supplemented with a mix of feed additives intended to promote fish growth performance and feed intake. The ECO diets presented lower production costs than the COM diet, whilst incorporating more sustainable ingredients. Gilthead seabream juveniles (±154.5 g initial body weight) were randomly assigned to triplicate tanks and fed the diets for 84 days. Fish fed the ECOSup diet attained a similar final body weight than fish fed the COM diet, significantly higher than fish fed the ECO diet. ECOSup fed fish presented significantly higher hepatosomatic index than COM fed fish, most likely due to higher hepatic glycogen reserves. The viscerosomatic index of ECOSup fed fish were significantly lower compared to COM fed fish, which is a positive achievement from a consumer's point of view. ECOSup diet exhibited similar nutrient digestibility than the COM diet. Moreover, feeding fish with the ECO diets resulted in lower faecal nitrogen losses when compared to COM fed fish. The results suggest that feeding gilthead seabream with an eco-friendly diet with a mix of feed additives such as the ECOSup diet, promoted growth and minimised nitrogen losses to the environment. Nutritional strategies that ultimately promote feed intake and diet utilisation are valuable tools that may help conditioning fish to sustain growth even under low temperatures.

## 1. Introduction

Gilthead seabream (*Sparus aurata*) is the most cultivated and economic relevant marine fish species in Southern Europe and in the Mediterranean region and its global production exceeded 228 000 tonnes in 2018 (FAO, 2020). Since the species is mainly produced in cages (Seginer, 2016), fish are constantly exposed to natural seasonal changes during the grow-out period. Fish are poikilothermic animals, therefore influenced by water temperature, which affects their feed intake, digestion, metabolism and ultimately growth (Jobling, 1994). Growth performance of gilthead seabream juveniles under farming conditions is optimal at 24 – 26°C (Hernández et al., 2003) and this species is particularly vulnerable to low temperatures. During Winter, seabream drastically reduces feed intake and may even stop feeding below 13 °C (Ibarz et al., 2003; Tort et al., 1998). In addition, seabream shows an inability to adapt to cold as they do not resume feeding even when kept at low temperature for prolonged periods and when water temperature rises do not immediately restore normal feed consumption (Ibarz et al., 2007a; 2007b; Tort et al., 2004). This cold-induced fasting, causes significant economic losses in fish production as it can last for several months, resulting in low body weight gain or even weight losses (Ibarz et al., 2010; Tort et al., 1998).

Feeding trials are usually conducted under optimum conditions. A limited number of studies have aimed to optimise fish feed formulations under sub-optimal conditions and even fewer studies have focused on formulating specific diets to promote growth of gilthead seabream at low temperature. In previous studies (Richard et al., 2016; Schrama et al., 2017; Silva et al., 2014), a high-quality diet containing high levels of marine-derived protein sources and nutritional supplements was offered to seabream juveniles through Winter and Spring, under natural temperature (mean values: 13°C in Winter and 18°C in Spring) and photoperiod conditions. The supplemented diet seemed to partially counteract the negative effects of exposure to low temperature on fish growth performance and to improve its nutritional and metabolic status through the seasons. However, the aforementioned experimental diets were neither economically viable nor environmentally sustainable as they relied on the inclusion of high levels of protein (~50% crude protein) and fishmeal (40%).

The economic viability and the environmental sustainability of the Aquaculture industry is closely entangled with the reduction of the dietary protein content as well as in the utilisation of more sustainable ingredients to produce aquafeeds. Although crude protein requirements are dependent upon protein digestibility and amino acid profile, dietary incorporation of 45% crude protein has proven to promote growth in gilthead seabream juveniles of 100-200 g (Lupatsch et al., 2003; Santinha et al., 1996). Protein is the most expensive nutrient in the diets (Rana et al., 2009) and therefore, reducing dietary protein inclusion has a positive economic impact, while may result in lower nitrogen excretion into the environment (Bureau and Hua, 2010; Teodósio et al., 2020 (Chapter II)]. One strategy to increase the Aquaculture sustainability is the replacement of marine-derived ingredients such as fishmeal, by ingredients that are eco-friendly but also highly digestible and able to sustain fish growth. Recent studies have pointed towards the use of poultry meal in seabream diets as a good alternative to fishmeal due to its protein content, digestibility and palatability (Davies et al., 2019; Fontinha et al., 2020; Sabbagh et al., 2019). Since it is a by-product of poultry processing plants and slaughterhouses, poultry meal is an environmentally sustainable and cost-effective alternative to fishmeal (Jedrejek et al., 2016).

Nutritional strategies may help conditioning fish to sustain growth even under adverse conditions. The incorporation of feed additives in aquafeeds offers interesting possibilities in fish nutrition (Encarnação, 2016), and might be added for several purposes. For instance, as a phagostimulant, betaine supplementation has increased feed intake in several fish species, such as European seabass, *Dicentrarchus labrax* (Dias et al., 1997), brown-marbled grouper, *Epinephelus fuscoguttatus* (Lim et al., 2015) and gibel carp, *Carassius gibelio* (Xue and Cui, 2001). Since plant protein sources are taurine-deficient, taurine is also frequently added to plant-based diets due to its role in lipid digestion, bile acid conjugation and antioxidant defence (Salze and Davis, 2015), and as an attractant and feed stimulant (Chatzifotis et al., 2009). Also, diet palatability may be enhanced through the inclusion of krill meal, as showed in Pacific white shrimp, *Litopenaeus vannamei* (Derby et al., 2016) and blue shrimp, *Litopenaeus stylirostris* (Suresh et al., 2011). This ingredient has the advantage of also being an excellent source of marine phospholipids (Saleh et al., 2013a; 2018). Phospholipids, from marine and plant origin, have proven to enhance fish growth by improving lipid digestion and/or absorption (Cahu et al., 2003; Saleh et al., 2013b; Tocher et al., 2008). All these nutritional strategies that may promote feed intake and diet utilisation are valuable when formulating diets to help fish cope under adverse conditions.

The present study aimed to evaluate fish growth performance, feed utilisation, nutrient apparent digestibility, and nitrogen outputs to the environment in fish fed experimental diets with lower protein content and environmentally sustainable. The experimental diets differed in the inclusion of feed additives which were incorporated to stimulate feed intake and enhance weight gain. The main goal of this study is to contribute to the optimisation of economically viable and environmentally sustainable diets that are suitable for gilthead seabream ongrowing production at low temperatures.

## 2. Material and Methods

### *2.1. Experimental diets*

Three isolipidic (crude fat: ~17.7% as fed) and isoenergetic (gross energy: ~20.9 MJ $kg^{-1}$ as fed) diets were formulated using practical ingredients (Table 6.1). A high protein diet (COM) was formulated to be similar to a commercial feed used for gilthead seabream juveniles, with fishmeal (27.5%) and soy ingredients (14%) as the main protein sources and a crude protein (CP) content of 44%. The other two diets, ECO and ECOSup, were formulated to reduce dietary fishmeal inclusion and protein content (42% CP). In these diets, poultry meal (15%), soy ingredients (11.5%) and fishmeal (10% for ECO and 7.5% for ECOSup) were used as protein sources. In addition, the ECOSup diet contained a mix of feed additives to potentially promote feed intake and feed utilisation, such as: betaine (1%), krill meal (5%), soy lecithin (1%), macroalgae mix (1%) and L-taurine (0.3%). All diets were supplemented with selected indispensable amino acids (IAA) and mono-calcium phosphate, whenever necessary, to fulfil the nutritional requirements of juvenile gilthead seabream. The ECO and ECOSup formulation costs were 86 and 95% relative to the COM diet. The costs of the experimental diets were calculated by the feed manufacturer, SPAROS Lda. (Olhão, Portugal).

Diets (pellet size 3 mm) were produced at SPAROS Lda. by extrusion by means of a pilot-scale twin-screw extruder (CLEXTRAL BC45; Clextral, France) with a screw diameter of 55.5 mm and temperature ranging from 105°C to 110°C. Upon extrusion, all batches of extruded feeds were dried in a vibrating fluid bed dryer (model DR100; TGC Extrusion, France). Following drying, pellets were allowed to cool at room temperature and subsequently the oil fraction was added under vacuum coating in a Pegasus vacuum mixer (PG-10VCLAB; DINNISEN, The Netherlands). Additionally, to measure the apparent digestibility of the diets by the indirect method, 5 kg of each diet was reground, chromic oxide was incorporated at 1% and the mixtures were dry-pelleted (screen diameter: 4.5 mm), using a steamless pelleting machine (CPM- 300; San Francisco, USA). Throughout the duration of the trial, experimental feeds were stored at room temperature, in a cool and aerated storage room. Proximate composition and amino acid analysis were performed for all experimental diets, as reported in Tables 6.1 and 6.2, respectively.

### *2.2. Zootechnical trials*

Experiments were carried out in compliance with the Guidelines of the European Union Council (Directive 2010/63/EU) and Portuguese legislation for the use of laboratory animals. Animal protocols were performed under Group-C licenses by the Direção Geral de Alimentação e Veterinária, Ministério da Agricultura, Florestas e Desenvolvimento Rural, Portugal.

Gilthead seabream juveniles (*Sparus aurata*) were obtained from Atlantik Fish Lda. (Castro Marim, Portugal) and the experiments were conducted at the Ramalhete Experimental Research Station of the Centre of Marine Sciences (CCMAR, Faro, Portugal). Upon arrival, fish were adapted to new conditions for about one month in a flow-through system with aeration, during which they were fed a commercial diet. Mean water temperature during the adaptation period was 13.0 ± 1.3°C.

### *2.2.1. Digestibility trial*

The apparent digestibility coefficients (ADC) of the dietary components were determined by the indirect method, using 1% chromic oxide as a dietary inert tracer, in nine homogeneous groups of gilthead seabream with a mean body weight of 153.5 ± 0.8 g. Triplicate groups of fish (9 fish per tank) were allocated to cylinder-conical 200 L tanks coupled with faeces collectors. Water average temperature was 14.2 ± 0.8°C, salinity 34.0 ± 0.8 psu and dissolved oxygen in water was 90.3 ± 3.9% of saturation. Fish were allowed to adapt to new conditions for one week before starting faeces collection. During the adaptation period fish were fed by hand to apparent satiety once a day one of the experimental diets and continue to do so throughout the trial. Tanks were thoroughly cleaned to remove any uneaten feed. Before feeding, faeces were collected daily for 12 days, left to settle and water was decanted. Faeces were frozen at –20°C and freeze-dried prior to analysis.

The apparent digestibility coefficients (ADC) of the dietary nutrients and energy were calculated as follows (Maynard et al., 1979):

$$\text{ADC (\%)} = 100 \times \left[1 - \frac{\text{dietary } Cr_2O_3 \text{ level}}{\text{faecal } Cr_2O_3 \text{ level}} \times \frac{\text{faecal nutrient or energy level}}{\text{dietary nutrient or energy level}}\right]$$

ADC of dry matter was calculated as:

$$\text{ADC (\%)} = 100 \times \left[1 - \frac{\text{dietary } Cr_2O_3 \text{ level}}{\text{faecal } Cr_2O_3 \text{ level}}\right]$$

*2.2.2. Growth trial*

Fish were reared in 500 L cylindrical tanks supplied with flow-through aerated seawater (temperature: 13.4 ± 2.1°C; salinity: 34.4 ± 0.8 psu; dissolved oxygen in water above 90% saturation) under natural photoperiod conditions (January to mid-April). Daily water temperature data is presented as Supplementary Figure 1. Homogeneous groups of seabream juveniles with a mean body weight of 154.5 ± 13.8 g were distributed in groups of 4 fish, into nine tanks at an initial density of 8.6 kg $m^{-3}$ (28 fish per tank). Five fish from the initial stock were sampled and stored at –20°C for subsequent analysis of whole-body composition. Each experimental diet was randomly assigned to triplicate tanks and tested for 84 days. Fish were fed by hand to apparent satiety once a day (10h00), except Sundays, avoiding feed losses and apparent feed intake was recorded. Mortality, water oxygen saturation and temperature were monitored daily. To monitor growth and feed utilisation, fish from each tank were bulk weighed under moderate anaesthesia every four weeks.

At the end of the trial, each tank was bulk weighed. Twenty fish from each tank were euthanised with a lethal dose of anaesthetic (1.5 mL $L^{-1}$ 2-phenoxyethanol; Sigma-Aldrich, Spain) and individually weighed. Total length of 10 fish per tank (*n* = 30 per treatment) was recorded to determine condition factor (K). Five fish from each tank were pooled and stored at –20°C until analysis of whole-body composition ($n$ = 3 pools per treatment). Liver and viscera weight of five fish per tank ($n$ = 15 per treatment) were recorded for calculation of hepatosomatic and viscerosomatic indexes and liver were stored at –20°C until protein and lipid analysis. Fish were fasted for 24 h before initial and final samplings.

### *2.3. Chemical analysis*

Chemical analysis followed standard procedures of the Association of Official Analytical Chemists (AOAC, 2006) and were done in duplicates. Before analysis, diets, faeces and pooled whole-body fish were finely ground. Diets and whole-body fish proximal composition was determined as follows: dry matter by drying the samples at 105°C for 24 h and ash content by incineration in a muffle furnace at 550°C for 6 h. Freeze-dried diets, whole-body fish and faeces samples were analysed for crude protein (N x 6.25) using a Leco nitrogen analyser (Model FP- 528; Leco Corporation, St. Joseph, USA); crude fat by petroleum ether extraction using a Soxtherm Multistat/SX PC (Gerhardt, Germany); gross energy by combustion in an adiabatic bomb calorimeter (Werke C2000; IKA, Staufen, Germany) calibrated with benzoic acid; phosphorus content by digestion at 230°C in a Kjeldatherm block digestion unit followed by digestion at 75°C in a water bath and absorbance determination at 820 nm (adapted from AFNOR V 04–406) and chromic oxide content was determined according to Bolin et al. (1952), after digestion with perchloric acid. Liver samples were pooled together by replicate tank, freeze-dried and analysed for protein content, as described above, and fat content according to Bligh and Dyer (1959).

Total amino acid profile from both faeces and diets were determined by ultra-high-performance liquid chromatography (UPLC) on a Waters Reversed-Phase Amino Acid Analysis System, using norvaline as an internal standard. Samples were pre-column derivatised with Waters AccQ Fluor Reagent (6-aminoquinolyl-N-hydroxysuccinimidyl carbamate) using AccQ Tag method (Waters, USA) after acid hydrolysis (HCl 6 M at 116°C for 48 h in nitrogen-flushed glass vials). Amino acids were identified by retention times of standard mixtures (Waters) and pure standards (Sigma-Aldrich). Instrument control, data acquisition and processing were achieved by the use of Waters Empower software.

*2.4. Calculations*

Key performance indicators were calculated as follows:

Weight gain (%) = 100 × wet weight gain × initial biomass$^{-1}$, where wet weight gain is: final biomass – initial biomass.

Thermal growth coefficient (TGC) = 100 × (FBW$^{1/3}$ – IBW$^{1/3}$) × DD$^{-1}$, where IBW and FBW are the initial and final body weight, respectively, and DD is the sum of degree.days for the experimental period.

Daily voluntary feed intake (VFI, % day$^{-1}$) = 100 × apparent feed intake × ABM$^{-1}$ × days$^{-1}$, where ABM is average body mass = (final biomass + initial biomass)/2.

Feed conversion ratio (FCR) = apparent feed intake × wet weight gain$^{-1}$.

Protein efficiency ratio (PER) = wet weight gain × crude protein intake$^{-1}$.

Condition factor (K) = 100 × body weight × total length$^{-3}$.

Hepatosomatic index (HSI %) = 100 × liver weight × body weight$^{-1}$.

Viscerosomatic index (VSI %) = 100 × viscera weight × body weight$^{-1}$.

Nitrogen (N) gain (mg N kg$^{-1}$ day$^{-1}$) = (final whole-body N content – initial whole-body N content) × ABM$^{-1}$ × days$^{-1}$.

Faecal N loss (mg N kg$^{-1}$ day$^{-1}$) = [N intake × (100 – N ADC %)] × ABM$^{-1}$ × days$^{-1}$.

Metabolic N loss (mg N kg$^{-1}$ day$^{-1}$) = N intake – (N gain + faecal N losses).

### 2.5. Statistical analysis

Data are presented as means ± standard deviation. Data expressed as a percentage were arcsine square root transformed previously to the statistical analysis (Ennos, 2012). All data were checked for normal distribution and homogeneity of variances. Differences among dietary treatments were identified by one-way analysis of variance (ANOVA) followed by Tukey's multiple-comparison test at $P < 0.05$ level of significance. Statistical analyses were performed using the open source software R version 4.0.1.

## 3. **Results**

### 3.1. Digestibility trial

Apparent digestibility coefficients (ADC) of nutrients and energy of experimental diets are presented in Table 6.3. Protein and fat digestibility were high in all diets and were not influenced ($p > 0.05$) by the dietary treatment. Phosphorus ADC values were significantly higher ($p < 0.05$) for ECOSup and COM diets than for ECO diet. Energy digestibility was significantly higher ($p < 0.05$) in diet ECOSup than in ECO diet and not significantly different from the COM diet ($p > 0.05$). Based on these results, calculated values of digestible protein to digestible energy (DP: DE) ratios were 21.9 for the COM diet and 21.2 for both ECO and ECOSup diets (Table 6.3). As for the amino acids, ADC values for the ECOSup and COM diets presented similar values (Table 6.4), with the exception of phenylalanine and tyrosine that presented significantly lower values in the former ($p < 0.05$). Amino acid digestibility of the ECO diet was in general lower than the ECOSup or COM diets.

### 3.2. Growth trial

Twenty-eight and 56 days after being fed the experimental diets, gilthead seabream juveniles presented similar growth performance parameters ($p > 0.05$; data not shown).

However, at the end of the growth trial (84 days), seabream fed the ECOSup and the COM diets presented similar ($p > 0.05$) final body weight of 191.4 ± 22.8 g and 193.7 ± 28.8 g, respectively, while ECO fed fish presented a significantly lower ($p < 0.05$) body weight (179.7 ± 29.2 g), from fish fed the COM diet (Table 6.5). It is worth noticing that in this period of low temperatures, all fish were able to increase initial body weight by 25%, 22% and 19% when fed the COM, ECOSup and ECO diets, respectively. Although TGC, VFI, FCR and PER were unaffected by the dietary treatments ($p > 0.05$), ECOSup and COM fed fish presented better performance indicators when compared with fish fed the ECO diet (Table 6.5). Fish condition factor (K) was similar in all dietary treatments, however fish fed the ECOSup diet presented significantly higher hepatosomatic and lower viscerosomatic indexes when compared to fish fed COM diet ($p < 0.05$; Table 6.5). During the trial, survival was high (~98%) and unaffected by the dietary treatments ($p > 0.05$).

At the end of the trial, whole-body composition was not significantly affected ($p > 0.05$) by the dietary treatments (Table 6.6). Moisture content was around 65%, protein content was higher than 16.5%, whole-body fat content presented mean values of approximately 12.5%, phosphorus ranged from 0.8 to 0.9% and energy content varied from 7.8 to 8.1 MJ $kg^{-1}$ (all values in wet weight basis). Dietary treatments did not affect ($p > 0.05$) hepatic protein and fat content (Table 6.6). Furthermore, fish nutrient and energy retention were not affected ($p > 0.05$) by the dietary treatments (results not shown).

### *3.3. Nitrogen balance*

Whole-body composition analysis combined with information on ADC of diets allowed the calculation of daily nitrogen balance (Figure 6.2). Daily nitrogen gain and metabolic losses were unaffected ($p > 0.05$) by the dietary treatments. Daily nitrogen gain varied from 72.7 ± 15.4 to 59.2 ± 22.8 mg N $kg^{-1}$ $day^{-1}$ and metabolic losses from 245.2 ± 23.2 to 203.9 ± 24.8 mg

N $kg^{-1}$ $day^{-1}$ for fish fed COM and ECO diets, respectively. ECOSup fed fish displayed intermediate values. However, feeding gilthead seabream juveniles with lower protein content diets ECO and ECOSup resulted in a significant reduction in nitrogen faecal losses. Fish fed the COM diet presented N faecal losses of 39.2 ± 1.3 mg N $kg^{-1}$ $day^{-1}$, while ECO and ECOSup fed fish lost 29.3 ± 2.7 and 24.6 ± 1.9 mg N $kg^{-1}$ $fish^{-1}$ respectively, to the environment.

## 4. Discussion

The present study aimed to evaluate growth performance and diet digestibility in gilthead seabream juveniles fed a diet similar to a commercial feed (COM) and two eco-friendly and less expensive feeds, ECO and ECOSup, under low temperatures (mean water temperature ~13°C). The ECO diets were formulated not only to include environmentally sustainable ingredients in cost-effective feeds, but also to sustain growth under low temperature conditions.

At the end of the growth trial, fish fed the ECOSup diet presented similar growth performance results than fish fed the commercial diet (COM), while ECO fed fish presented significantly lower body weight than fish fed the COM diet. Nutritional studies performed under low water temperature conditions are scarce. However, in a study that evaluated gilthead seabream juveniles (initial weight ± 87 g) growth performance through Winter (mean water temperature ~13°C), individual fish gained 0.08 g per day when fed a control diet (commercial diet, 47% CP as fed basis) and 0.13 g per day when fish were fed a Winter diet (48% CP as fed basis) supplemented with a mix of additives (Silva et al., 2014). Additionally, TGC was improved from 0.01 to 0.02 and FCR from 4.5 to 2.4 when fish were fed the Winter diet. In another study performed at 14°C, after being fed three isonitrogenous (47% CP on dry matter basis) diets that differed in lipid content (14%, 16% and 18%) for 50 days, seabream (initial weight ± 145 g) exhibited similar growth performance indicators for all dietary treatments: mean weight gain was 0.41 g $fish^{-1}$ $day^{-1}$ and FCR ranged from 2.5 to 2.6 (Sánchez-Nuño et al., 2018). In the present study, growth performance parameters were better than previous published experiments under low water temperature: mean weight gain per fish was 0.44 g per day, TGC presented an average value of 0.03 for all dietary treatments and FCR values varied from 1.8 to 2.1. The current results suggest that under low temperature conditions, the ECOSup diet is a valuable alternative to the COM diet in terms of growth and feed utilisation since feeding fish with these diets resulted in similar fish body weight, feed conversion and protein efficiency ratios.

Feeding fish with lower protein and fishmeal content diets (ECO and ECOSup) resulted in similar whole-body composition to fish fed a commercial diet (COM). Whole-body nutrient and energy contents were marginally below or in line with published data for seabream juveniles (Dias et al., 2009; Kissil and Lupatsch, 2004). Fish ingest feed according to their energy demands and at low temperature seabream reduce metabolic activity, lowering their energy requirements (Ibarz et al., 2003). However, lower feed intake may cause a deficit in energy intake that will lead to the utilisation of body energy reserves, preventing fat deposition as perivisceral fat (Ibarz et al., 2005). Fish fed the ECOSup diet presented a significantly lower viscerosomatic index than COM fed fish. No differences were found among treatments concerning whole-body or liver fat content, therefore, it is reasonable to assume that only visceral fat differs between ECOSup and COM fed fish. Fish containing low visceral fat are more likely to be accepted from a consumer’s point of view.

Fish fed the ECOSup diet exhibited a significantly higher hepatosomatic index (HSI) when compared to fish fed the COM diet. Generally, a higher HSI is associated with an increase in fat content in liver (Rueda-Jasso et al., 2004). However, this does not seem to be the case in ECOSup fed fish as there are no significant effects of the dietary treatments in hepatic fat content. Contrary to our results, previous studies have documented fat accumulation in liver of fish subjected to low temperature or thermal shifts (Gallardo et al., 2003; Ibarz et al., 2007a;

Ibarz et al., 2005). Furthermore, given the fact that hepatic protein content was also similar among treatments, the differences found in HSI among treatments might be due to changes in energy metabolism and storage of glycogen in the liver. This hypothesis is in agreement with a previous study that used FT-IR spectroscopy to understand how dietary factors affected liver metabolic content in seabream exposed to seasonal temperature variations (Silva et al., 2014). In that study, at the end of Winter, fish fed a commercial diet showed a lower carbohydrate storage than fish fed a Winter feed. The current results suggest that feeding seabream with the ECOSup diet helps to mitigate the negative effects of low water temperature. Higher energy reserves imply improved fish nutritional status and may be an advantage when optimal growth conditions occur or during periods of low feed intake.

The COM diet had a higher inclusion of fishmeal (27.5%) than the ECO diets. Both ECO diets contained 15% of poultry meal and included fishmeal at 10% (ECO) and 7.5% (ECOSup). Fishmeal has high palatability and is an excellent source of highly digestible protein for most aquatic species (Turchini et al., 2019). It has been considered the gold standard protein source in aquafeeds, especially when feeding carnivorous fish such as the gilthead seabream. However, the dependency of the aquafeed industry on fishmeal has been an environmental concern and a limitation for the continuous growth of the activity; thus, great efforts have been made on finding alternative protein sources (Matos et al., 2017). Due to the European Union (EU) ban in 2001 on the use of terrestrial animal proteins in animal feeds (European Parliament, 2001), research efforts have been focused on the use of vegetable protein sources to replace fishmeal in aquafeeds (Gatlin et al., 2007). However, in 2013 the use of non-ruminant processed animal proteins (PAP) such as poultry meal, has been reinstated in the EU (European Commission, 2013). Poultry meal is made from recycling by-products of slaughterhouses and processing plants, therefore, using poultry meal as a feed ingredient brings value to what would otherwise be perceived as waste material. Furthermore, poultry meal is a highly abundant commodity and is available in markets worldwide (Tacon et al., 2011). Taking all of this into consideration, poultry meal is considered as an environmentally sustainable protein source, with high digestibility and palatability (Oliva-Teles et al., 2015). Previous studies showed that it is possible to replace 50% of fishmeal with poultry meal in seabream diets without compromising growth performance and feed efficiency (Davies et al., 2019; Karapanagiotidis et al., 2019). Other studies in seabream juveniles propose that fishmeal substitution by poultry meal can be up to 83% or even 100% without hindering growth or diet utilisation (Fontinha et al., 2020; Sabbagh et al., 2019). In the present study, no detrimental effects on growth performance and diet utilisation were observed in fish fed the ECOSup diet under low temperature conditions, while fish fed the ECO diet presented lower body weight compared to fish fed the commercial diet (COM). This indicates that poultry meal may be successfully used to replace high dietary levels of fishmeal at low water temperature conditions, if supplemented with an adequate mix of feed additives.

The ECOSup diet contained a mix of feed additives to potentially promote feed intake and utilisation, such as: betaine, taurine, krill meal, soy lecithin and macroalgae mix. Although previous studies showed higher feed consumption by adding betaine (Kolkovski et al., 1997; Xue et al., 2004) and taurine (Chatzifotis et al., 2009) to the diets, in the present study feed intake was not affected by the inclusion of feed additives. Further investigation is needed to fine-tune the supplementation levels of these compounds as feed additives for the different ongrowing stages as well as rearing conditions. Krill meal and soy lecithin, as sources of phospholipids, have been shown to increase lipid digestion and absorption in seabream larvae (Saleh et al., 2013a; 2013b). In fact, fish fed the ECOSup diet had a lower viscerosomatic index than fish fed the COM diet, suggesting that the supplementation of ECOSup diet resulted in a lower fat deposition due to better utilisation of the dietary lipids. Although the improvement in growth performance and nutrient digestibility cannot be explained by a single additive, the combination of these compounds positively affected gilthead seabream fed the ECOSup diet.

Evaluation of feed digestibility is crucial to maximise nutrient utilisation and minimise aquaculture environmental impact when formulating new diets. In general, seabream fed the experimental diets revealed a high capability to digest nutrients and energy. Digestibility data obtained in the current study is within the range of values reported for seabream fed diets with similar fishmeal and plant protein inclusion levels (Aragão et al., 2020; Dias et al., 2009). The ECOSup diet presented the highest digestibility values for all nutrients, although only differing significantly from the ECO diet for phosphorus and energy. Regarding amino acid digestibility, the ECOSup diet showed results similar to the COM diet, significantly higher than the ones obtained for the ECO diet for most of the amino acids. The higher digestibility of the ECOSup diet implies a greater availability of energy and nutrients to the fish. This is likely to be responsible for the improved growth performance observed in the feeding trial by gilthead seabream fed the ECOSup diet when compared to the ECO fed fish, reinforcing the positive effects of the mix of feed additives to a low protein and low fishmeal diet.

Nitrogen balance calculations for fish fed the different diets revealed that nitrogen gain and metabolic losses were unaffected by the dietary treatments. However, fish fed the lower protein diets ECOSup and ECO, presented significantly lower nitrogen faecal losses than fish fed the COM diet. Fish fed the ECOSup diet lost 7.7% of nitrogen via faeces, a low value especially compared with fish fed the COM diet that lost 11% of the nitrogen intake. Although no significant differences were observed in protein digestibility among treatments, the ECOSup diet presented the highest protein digestibility, which explains lower nitrogen faecal losses and therefore, the release of lower amounts of nitrogen into the aquatic environment. It is vital that environmentally sustainable diets are at optimal requirements for the fish and incorporate highly digestible ingredients so that nutrients losses are reduced (Matos et al., 2017). The reduction of the dietary protein content combined with a higher protein digestibility causes a major positive impact in the reduction of nitrogen release in the environment. The present results support the fact that changing from a COM diet to lower protein diets such as the ECO diets, will help the aquaculture sector to achieve long- term environmental sustainability.

Formulating diets that reduce the inclusion of dietary protein as well as fishmeal by replacing it with more sustainable and economic ingredients, is a viable strategy to lower aquaculture production costs and minimise its environmental impact. The reduction of the protein content in aquafeeds has a major impact in the economic viability of the sector. Protein is the major cost associated with feeds and feeds are the main costs of production. The ECO diets presented a lower production cost compared to the COM diet; the ECO diet was 86% of the COM diet while the ECOSup was 95%. This difference between the ECO diets was due to the inclusion of a mix of feed additives in the ECOSup diet that aimed to stimulate feed intake and enhance weight gain. In fact, the ECOSup diet, but not the ECO, resulted in similar fish growth when compared to the commercial diet COM, while reducing nitrogen faecal losses.

## 5. Conclusions

This study supports the concept that nutrition is a powerful tool to tailor-made diets that will help fish coping with exposure to low water temperatures. Moreover, supplementing a low protein and fishmeal diet with a mix of feed additives as is the case of the ECOSup diet, promoted growth and minimised nitrogen losses to the environment. The current findings suggest that feeding gilthead seabream juveniles with the ECOSup diet may have a major positive impact in the environmental sustainability of the aquaculture sector, while sustaining growth in a critical period such as Winter.

**Acknowledgements**

RT acknowledges financial support by Evonik Nutrition & Care GmbH (Germany).

**Funding**

This work was supported by European Union's Horizon 2020 research and innovation programme under Project MedAID (GA no 727315) and by the Portuguese Foundation for Science and Technology (Ministry of Science and Higher Education, Portugal) through project UIDB/04326/2020 to CCMAR, contract DL 57/2016/CP1361/CT0033 to CA. The views expressed in this work are the sole responsibility of the authors and do not necessary reflect the views of the European Commission.

**Figures**

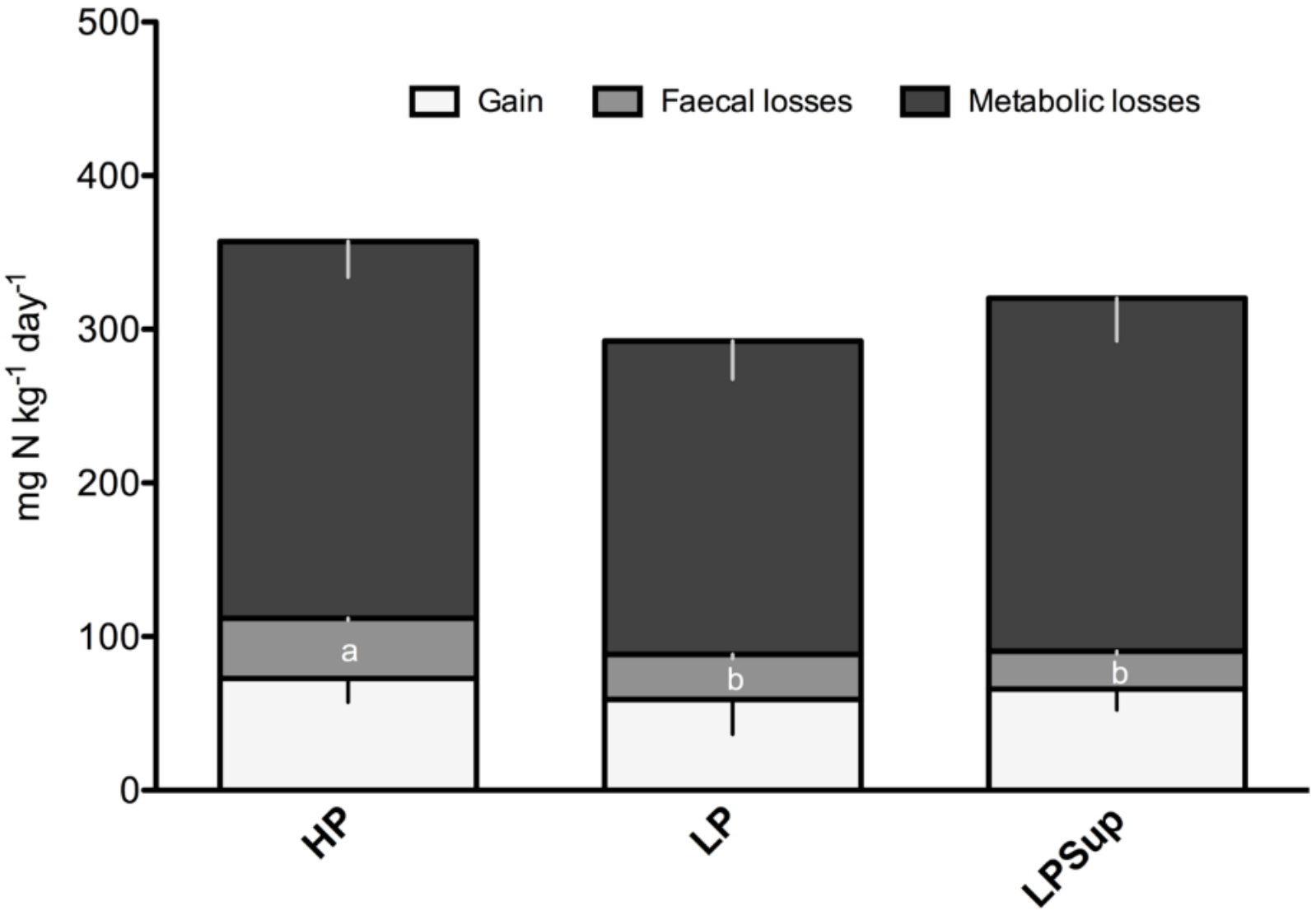

**Figure 1** Daily nitrogen (N) balance in gilthead seabream juveniles fed the experimental diets for 84 days. Values are presented as means ± standard deviation ($n$ = 3). Different superscripts within bars indicate significant differences ($p < 0.05$) among diets. Absence of superscripts indicates no significant differences ($p > 0.05$).

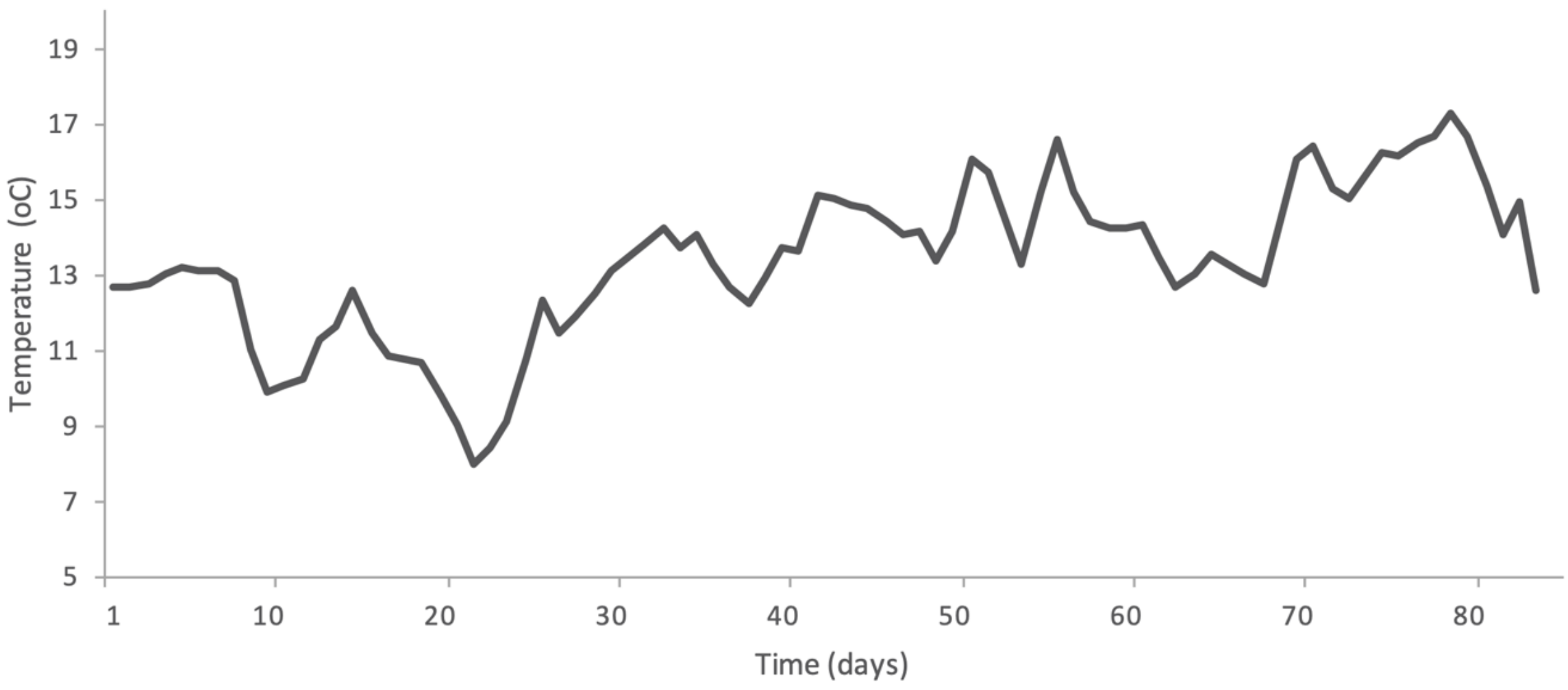

**Supplementary Figure 1** Water temperature profile throughout the growth trial.

**Table 1** Formulation and proximate composition of experimental diets.

| *Ingredients (%)* | **COM** | **ECO** | **ECOSup** |
|---|---|---|---|
| Fishmeal [a] | 27.50 | 10.00 | 7.50 |
| Fish soluble protein concentrate [b] | 2.50 | 2.50 | 2.50 |
| Squid meal [c] | 2.50 | 2.50 | 2.50 |
| Krill meal [d] | 2.50 | 2.50 | 5.00 |
| Poultry meal [e] | | 15.00 | 15.00 |
| Soy protein concentrate [f] | 10.00 | 7.50 | 7.50 |
| Wheat gluten [g] | 4.00 | 4.00 | 4.00 |
| Corn gluten [h] | 7.50 | 6.00 | 6.30 |
| Soybean meal [i] | 4.00 | 4.00 | 4.00 |
| Wheat meal [j] | 16.85 | 22.65 | 20.45 |
| Faba beans (low tannins) [k] | 6.00 | 6.00 | 6.00 |
| Sardine oil [l] | 10.22 | 10.15 | 9.52 |
| Rapeseed oil [m] | 4.38 | 4.35 | 4.08 |
| Soy lecithin [n] | | | 1.00 |
| Vitamin and Mineral Premix [o] | 1.00 | 1.00 | 1.00 |
| Lutavit C35 and E50 [p] | 0.05 | 0.05 | 0.05 |
| Betaine HCl [q] | 0.50 | 0.50 | 1.00 |
| Macroalgae mix [r] | | | 1.00 |
| Antioxidant powder [s] | 0.20 | 0.20 | 0.20 |
| Sodium propionate [t] | 0.10 | 0.10 | 0.10 |
| Mono-calcium phosphate [u] | | 0.40 | 0.40 |
| L-Lysine [v] | | 0.20 | 0.20 |
| L-Tryptophan [w] | | 0.10 | 0.10 |
| DL-Methionine [x] | 0.20 | 0.30 | 0.30 |
| L-Taurine [y] | | | 0.30 |
| | | | |
| *Proximate composition (% as fed)* | | | |
| Dry matter | 92.9 | 94.0 | 94.9 |
| Ash | 7.4 | 6.9 | 7.2 |
| Crude protein | 44.2 | 42.0 | 42.0 |
| Crude fat | 18.0 | 17.6 | 17.6 |
| Total phosphorus | 1.0 | 1.1 | 1.0 |
| Gross energy (MJ $kg^{-1}$) | 20.9 | 20.9 | 20.8 |
| CP/GE | 21.1 | 20.1 | 20.2 |

CP/GE: crude protein to gross energy ratio.

All values are reported as means of duplicate analysis.

[a] Super Prime: 66.3% crude protein (CP), 11.5% crude fat (CF); Pesquera Diamante, Peru.

[b] CPSP 90: 84% CP, 12% CF; Sopropêche, France.

[c] Super prime without guts: 84% CP, 4.7% CF; Sopropêche, Spain.

[d] Krill meal: 61.1% CP, 17.4% CF; Aker Biomarine, Norway.

[e] Poultry meal 65: 67% CP, 12% CF; Sonac, The Netherlands.

[f] Soycomil P: 63% CP, 8% CF; ADM, The Netherlands.

[g] VITAL: 80% CP, 7.5% CF; Roquette Frères, France.

[h] Corn gluten meal: 61% CP, 6% CF; COPAM, Portugal.

[i] Solvent extracted dehulled soybean meal: 47% CP, 2.6% CF; CARGILL, Spain.

[j] Wheat meal: 10% CP, 1.2% CF; Casa Lanchinha, Portugal.

[k] Faba beans low tannins: 28% CP, 1.2% CF; Casa Lanchinha, Portugal.

[l] Sopropêche, France.

[m] J.C. Coimbra Lda., Portugal.

[n] Lecico P700IPM; LECICO GmbH, Germany.

[o] INVIVONSA Portugal AS, Portugal: Vitamins (IU or mg $kg^{-1}$ diet): DL-alpha tocoferol acetate, 100 mg; sodium menadione bisulphate, 25 mg; retinyl acetate, 20000 IU; DL-cholecalciferol, 2000 IU; thiamin, 30 mg; riboflavin, 30mg; pyridoxine, 20 mg; cyanocobalamin, 0.1 mg; nicotin acid, 200 mg; folic acid, 15 mg; ascorbic acid, 500 mg; inositol, 500 mg; biotin, 3 mg; calcium panthotenate, 100 mg; choline chloride, 1000 mg; betaine, 500 mg. Minerals (g or mg $kg^{-1}$ diet): copper sulphate, 9 mg; ferric sulphate, 6 mg; potassium iodide, 0.5 mg; manganese oxide, 9.6 mg; sodium selenite, 0.01 mg; zinc sulfate, 7.5 mg; sodium chloride, 400 mg; excipient wheat middling's.

[p] BASF, Germany.

[q] Beta-Key 95%, ORFFA, The Netherlands.

[r] Macroalgae mix: 11% CP, 0.6% CF; Ocean Harvest, Ireland.

[s] Paramega PX, KEMIN EUROPE NV, Belgium.

[t] Disproquímica, Portugal.

[u] MCP: 22% P, 18% Ca, Fosfitalia, Italy.

[v] Biolys: L-lysine sulphate, 54.6% lysine; EVONIK Nutrition & Care GmbH, Germany.

[w] L-Tryptophan: 98%; EVONIK Nutrition & Care GmbH, Germany.

[x] DL-Methionine: 99%; EVONIK Nutrition & Care GmbH, Germany.

[y] L-Taurine: 98%; ORFFA, The Netherlands.

**Table 2** Amino acid composition of experimental diets.

| Amino acids *(mg AA g$^{-1}$ as fed)* | **COM** | **ECO** | **ECOSup** |
|---|---|---|---|
| Arginine | 34.2 | 28.3 | 30.8 |
| Histidine | 9.6 | 9.6 | 9.7 |
| Lysine | 26.8 | 24.8 | 29.0 |
| Threonine | 16.6 | 15.1 | 15.5 |
| Isoleucine | 20.6 | 18.3 | 19.1 |
| Leucine | 34.1 | 31.8 | 28.0 |
| Valine | 22.2 | 19.7 | 20.9 |
| Methionine | 14.5 | 14.2 | 12.0 |
| Phenylalanine | 23.0 | 19.3 | 18.6 |
| Cystine | 2.9 | 3.2 | 2.3 |
| Tyrosine | 18.8 | 18.3 | 15.3 |
| Aspartic acid + Asparagine | 34.6 | 32.7 | 35.1 |
| Glutamic acid + Glutamine | 67.8 | 66.4 | 66.5 |
| Alanine | 23.6 | 20.2 | 22.4 |
| Glycine | 30.3 | 23.2 | 27.7 |
| Proline | 30.0 | 25.8 | 26.0 |
| Serine | 17.4 | 15.5 | 15.6 |
| Taurine | 2.3 | 2.3 | 2.3 |

All values are reported as mean of duplicate analysis.

**Table 3** Apparent digestibility coefficients (ADC) of nutrients and energy of experimental diets.

| *ADC (%)* | **COM** | | **ECOSup** |
|---|---|---|---|
| Dry matter | 67.7 ± 1.3 | 8 | 69.6 ± 1.9 |
| Protein | 89.0 ± 1.5 | 2 | 92.3 ± 1.1 |
| Fat | 95.4 ± 0.8 | 4 | 96.2 ± 0.7 |
| Phosphorus | 54.6 ± 1.6$^{a}$ | 9$^{b}$ | 56.3 ± 2.2$^{a}$ |
| Energy | 85.8 ± 0.9$^{ab}$ | 3$^{b}$ | 88.1 ± 0.7$^{a}$ |
| | | | |
| DP/DE ratio | 21.9 | | 21.2 |

Values are presented as means ± standard deviation ($n$ = 3). Different superscripts (a, b) within the same row indicate significant differences ($p$ < 0.05) among diets. Absence of superscripts indicates no significant differences.

Abbreviations: DP, digestible protein; DE, digestible energy.

**Table 4** Apparent digestibility coefficients (ADC) of amino acids of experimental diets.

| *ADC (%)* | **COM** | | **ECOSup** |
|---|---|---|---|
| Arginine | $85.6 \pm 0.7^{a}$ | $4^{b}$ | $83.1 \pm 2.3^{a}$ |
| Histidine | $88.8 \pm 0.8$ | 5 | $88.6 \pm 1.4$ |
| Lysine | $88.5 \pm 0.5^{ab}$ | $8^{b}$ | $89.4 \pm 1.4^{a}$ |
| Threonine | $87.5 \pm 0.4^{a}$ | $2^{b}$ | $85.8 \pm 1.4^{ab}$ |
| Isoleucine | $91.7 \pm 0.3^{a}$ | $7^{b}$ | $91.0 \pm 1.1^{ab}$ |
| Leucine | $91.2 \pm 0.2$ | 6 | $89.1 \pm 1.4$ |
| Valine | $91.3 \pm 0.3$ | 6 | $91.0 \pm 1.2$ |
| Methionine | $91.3 \pm 0.5$ | 6 | $89.3 \pm 1.2$ |
| Phenylalanine | $91.9 \pm 0.3^{a}$ | $7^{b}$ | $89.6 \pm 1.4^{b}$ |
| Cystine | $88.6 \pm 0.7$ | 8 | $85.9 \pm 2.2$ |
| Tyrosine | $93.6 \pm 0.3^{a}$ | $6^{ab}$ | $92.1 \pm 0.9^{b}$ |
| Aspartic Acid + Asparagine | $74.8 \pm 1.5$ | 0 | $75.5 \pm 3.4$ |
| Glutamic acid + Glutamine | $90.5 \pm 0.5^{a}$ | $3^{b}$ | $88.8 \pm 1.3^{ab}$ |
| Alanine | $91.0 \pm 0.2^{a}$ | $6^{b}$ | $90.6 \pm 1.1^{a}$ |
| Glycine | $95.2 \pm 0.2^{a}$ | $4^{b}$ | $94.8 \pm 0.7^{a}$ |
| Proline | $91.6 \pm 0.3^{a}$ | $8^{b}$ | $89.7 \pm 1.2^{ab}$ |
| Serine | $86.1 \pm 0.3$ | 2 | $84.6 \pm 2.2$ |
| Taurine | $85.2 \pm 0.7$ | 5 | $87.3 \pm 1.3$ |

Values are presented as means ± standard deviation (n = 3). Different superscripts (a, b) within the same row indicate significant differences ($p < 0.05$) among diets. Absence of superscripts indicates no significant differences.

**Table 5** Growth performance and somatic indexes of gilthead seabream juveniles fed the experimental diets for 84 days.

| | **COM** | | **ECOSup** |
|---|---|---|---|
| FBW (g) | 193.7 ± 28.8[a] | 29.2[b] | 191.4 ± 22.8[ab] |
| WG (%) | 24.6 ± 3.0 | 6 | 22.3 ± 3.0 |
| TGC | 0.04 ± 0.01 | 01 | 0.04 ± 0.01 |
| VFI (% day$^{-1}$) | 0.51 ± 0.02 | 04 | 0.48 ± 0.04 |
| FCR | 1.8 ± 0.1 | | 1.9 ± 0.2 |
| PER | 1.3 ± 0.1 | | 1.3 ± 0.1 |
| K | 1.6 ± 0.1 | | 1.6 ± 0.1 |
| HSI (%) | 2.2± 0.3[b] | [ab] | 2.5 ± 0.4[a] |
| VSI (%) | 5.1 ± 0.7[a] | [ab] | 4.0 ± 0.6[b] |

Initial body weight = 154.5 ± 13.8 g for all dietary treatments (*n* = 252).

Values are presented as means ± standard deviation (*n* = 60 for FBW; *n* = 30 for K; *n* = 15 for HSI and VSI; *n* = 3 for the remaining parameters). Different superscripts (a, b) within the same row indicate significant differences ($p < 0.05$) among diets. Absence of superscripts indicates no significant differences.

Abbreviations: FBW, final body weight; WG, weight gain; TGC, thermal growth coefficient; VFI, daily voluntary feed intake; FCR, feed conversion ratio; PER, protein efficiency ratio; K, condition factor; HSI, hepatosomatic index; VSI, viscerosomatic index.

**Table 6** Whole-body and liver composition of gilthead seabream juveniles fed the experimental diets for 84 days.

| *Body composition (% wet weight)* | **COM** | | **ECOSup** |
|---|---|---|---|
| Moisture | 66.0 ± 0.3 | 5 | 65.1 ± 1.2 |
| Ash | 3.3 ± 0.5 | | 4.0 ± 0.8 |
| Protein | 16.6 ± 0.4 | 8 | 16.5 ± 0.3 |
| Fat | 12.3 ± 0.9 | 7 | 12.8 ± 0.9 |
| Phosphorus | 0.8 ± 0.1 | | 0.9 ± 0.1 |
| Energy (MJ $kg^{-1}$) | 7.8 ± 0.1 | | 8.1 ± 0.4 |
| | | | |
| *Liver composition (% dry weight)* | | | |
| Protein | 35.6 ± 3.0 | 2 | 32.1 ± 1.6 |
| Fat | 29.1 ± 3.4 | 3 | 22.1 ± 4.8 |

Initial body composition (% wet weight): moisture = 64.9 %; ash = 3.4 %; protein = 16.4 %; fat = 12.9 %; phosphorus = 0.8 %; energy = 8.2 MJ $kg^{-1}$.

Values are presented as means ± standard deviation ($n = 3$). Absence of superscripts indicates no significant differences ($p > 0.05$) among diets.